\documentclass[12pt]{iopart}
\usepackage{iopams}
\usepackage{graphicx}
\usepackage{epsfig}
\include{graphics}
\begin{document}

\title[Diverging PDF of flat-top solitons in an extended Korteweg-de Vries equation]{Numerical study on diverging probability density function of flat-top solitons in an extended Korteweg-de Vries equation}

\author{Yeojin Chung}

\address{Department of Mathematics, Southern Methodist University, 
Dallas, Texas 75275, USA}

\ead{ychung@smu.edu}
\date{\today}

\begin{abstract}
We consider an extended Korteweg-de Vries (eKdV) equation, the usual Korteweg-de Vries equation with inclusion of an additional cubic nonlinearity. We investigate the statistical behaviour of flat-top solitary waves described by an eKdV equation in the presence of weak dissipative disorder in the linear growth/damping term. With the weak disorder in the system, the amplitude of solitary wave randomly fluctuates during evolution. We demonstrate numerically that the probability density function of a solitary wave parameter $\kappa$ which characterizes the soliton amplitude exhibits loglognormal divergence near the maximum possible $\kappa$ value.

\end{abstract}
\maketitle

\section{Introduction}
It is well known that the Korteweg-de Vries (KdV) equation governs the propagation of shallow water waves of moderately small amplitude, where a balance between quadratic nonlinearity and linear dispersion results in soliton solutions. For the waves of large amplitudes or under certain circumstances in stratified fluids, however, it was found that an additional cubic nonlinearity becomes crucial, leading to the extended KdV (eKdV) equation \cite{grimshaw01,grimshaw02,ablowitz}. The eKdV equation, also known as the Gardner equation, appears as a governing equation for long interfacial waves in two-layer system \cite{ablowitz,ky78,kb81} as well as for oceanic stratification in shear flow \cite{grimshaw97,hpt97,mb98,gpt99}.

Both the KdV and eKdV equations are exactly integrable. The integrability induces that the eKdV equation possesses conventional soliton solutions of small amplitudes similar to those of the KdV equation. In addition, this extended equation exhibits a remarkable feature that distinguishes itself from the KdV equation, namely, the emergence of large amplitude wide solitons, called flat-top solitons. 
Such wide solitons also appear as solutions for other related nonlinear evolution equations led by a balance between dispersion and nonlinearity, such as high-order Nonlinear Schr\"odinger equations (NLS) or cubic-quintic complex Ginzburg-Landau equation. Consequently, flat-top solitary waves draw much attention in various areas of physics including fluid mechanics \cite{grimshaw01,gpps,so98,hpt01,js01}, nonlinear optics \cite{gagnon,soto,pt96}, and plasma physics \cite{zh92,tri}. 

Recently, we investigated the effects of weak dissipative disorder on flat-top solitary waves in cubic-quintic nonlinear Schr\"odinger equation (CQNLSE) and the derivative CQNLSE (DCQNLSE) \cite{avner05,avner09}. In particular, two most common types of disorder, disorders in the linear and cubic nonlinear gain/loss coefficients, are considered, which lead to the random variation of the solitary wave parameters including amplitude and group velocity. In this study, we showed numerically and analytically that the probability density function (PDF) of the soliton amplitude exhibits loglognormal divergence near the maximum possible amplitude value. This phenomenon stems from the fact that the solitary wave obtains a typical table-top shape when its amplitude approaches the maximum possible value. Thus, our anticipation was that the loglognormal divergence of amplitude can be generally associated with emergence of flat-top solitary waves. In the current paper, we corroborate this generality by showing that a perturbed eKdV equation whose structure differs from the CQNLSE and DCQNLSE, demonstrates similar statistical behaviour, namely, loglognormal divergence. 

The eKdV equation is not of the NLS type which was considered in our previous studies. On the other hand, it is ubiquitous and belongs to one of the families of integrable nonlinear partial differential equations. Therefore, it is important to examine if the loglognormal divergence of solitary wave parameters can be also found in a perturbed eKdV equation in the presence of weak disorder. In particular, we focus on the case when the dissipative disorder appears in the linear growth/damping term and its intensity is weak so that the solitary waves can evolve without severe distortion. This type of disorder can emerge quite commonly in systems that involve nonlinear wave equations. Indeed, for the case of water waves, as the depth of channel gradually increases or decreases, the evolution of waves can be described by a perturbed KdV equation, where a linear term proportional to the wave envelope is incorporated \cite{newell, miles81, miles79, KN80}. In the context of nonlinear optics, the random variations in the gain of amplifiers which are positioned to compensate the loss can lead to disorder in the linear gain coefficient \cite{kh83}. Such disorder also appears in massive multichannel transmission systems due to the interplay of Raman cross talk and bit pattern randomness \cite{p04,cp08}. 

Considering a perturbed eKdV equation where a random disorder appears in the linear growth/damping term, the amplitude of the solitary wave undergoes random fluctuations during evolution. Thus, we conduct Monte Carlo simulation to achieve the PDF of a parameter characterizing the soliton amplitude and verify its loglognormal divergence. This finding in turn, concludes that the loglognormal divergence of the amplitude PDF found in \cite{avner05,avner09} is not restricted to the solitary waves of NLS type equations. We also note that theoretical analysis of the perturbed KdV type equations is an extremely challenging task due to the substantial effects of radiation. More specifically, the linear perturbation induces a shelf consisting of radiative modes directly behind the solitary wave. While the shelf has a slowly varying small amplitude, its range extends with time, which varies at the rate of an order one \cite{ablowitz,newell,lr73}. 
 This phenomenon brings most of difficulties associated with the theoretical analysis of perturbed KdV type equations, and many questions still remain open despite various theoretical methods available \cite{newell,KN80,kn78,km78}. Although we expect that the adiabatic perturbation technique employed for the models in our previous study \cite{avner09} can be an appropriate tool to deal with the underlying problem, it calls for more extensive perturbative calculations in order to incorporate the full impact of shelf. We therefore defer the complete theoretical analysis to a future publication.

The material in this paper is organized as follows. In Sec. 2, we briefly describe the evolution of solitary waves of an eKdV equation in the presence of disorder in the linear growth/damping. In Sec. 3, we present the results of direct numerical simulations. Finally, in Sec. 4, we summarize our main results.

\section{Extended Korteweg-de Vries equation with disorder in the linear growth/damping coefficient}

We consider the evolution of solitary waves described by an eKdV equation with disorder in the linear growth/damping coefficient,
\begin{equation}
\partial_t u + 6u(1-\epsilon_n u)\partial_z u + \partial_z^3u = \epsilon \xi(t)u. \label{ekdv}
\end{equation}
In the context of internal waves, $u$ represents the amplitude of the wave (or the interfacial displacement), $z$ is the horizontal coordinate, and $t$ is time. The right hand side term $\epsilon \xi(t)u$ is responsible for the disorder effects of the linear growth/damping, and $\epsilon_n$ is the cubic nonlinear coefficient. We assume that the disorder $\xi(t)$ is zero in average and short correlated in time, i.e.,
\begin{equation}
\langle \xi(t) \rangle = 0,\;\;\langle \xi(t)\xi(t')\rangle = D\delta(t-t'),
\end{equation}
where $D$ is the disorder intensity. 

\begin{figure}
\begin{center}
\includegraphics[scale=0.5]{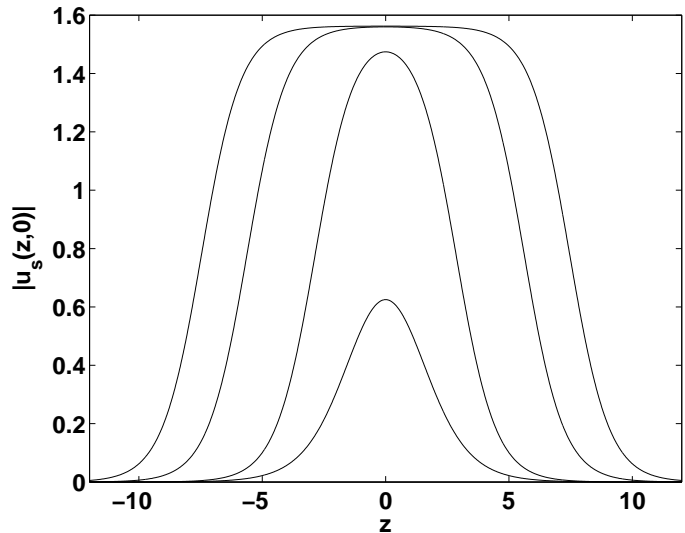}
\caption{Solitary wave solutions $u_s(z,t)$ at $t=0$ for different $\kappa$ values. From the lowest to the highest amplitude solutions, each solution corresponds to $\kappa=0.5,0.624,0.624\,999,0.624\,999\,99$.}
\label{soliton_shape}
\end{center}
\end{figure}

When $\epsilon=0$, we obtain an unperturbed eKdV equation whose soliton solutions are given by 
\begin{equation}
u_s(z,t) = \frac{4\kappa^2}{(1-\kappa^2/\kappa_m^2)^{1/2}\cosh(2x)+1}, \label{soliton}
\end{equation}
where $x=\kappa(z-4\kappa^2t)$, $\kappa_m = 0.5/\sqrt{\epsilon_n}$. Using the solution form (\ref{soliton}), we find that the parameter $\kappa$ characterizes the soliton amplitude and group velocity with the relation,
\begin{equation}
\kappa = \sqrt{\eta/2-\eta^2/(16\kappa_m^2)},\label{kappa}
\end{equation}
where $\eta$ represents the soliton amplitude.
These solitons are limited in amplitude and speed, namely, as $\kappa$ becomes close to its maximum possible value $\kappa_m$, the soliton forms the flat-top shape and the limiting flat-top soliton corresponds to the maximum amplitude and speed. Figure \ref{soliton_shape} illustrates the solitary wave solutions (\ref{soliton}), for different $\kappa$ values that range from $0.5$ to $0.624\,999\,99$, where we take $\epsilon_n=0.64$ corresponding to $\kappa_m = 0.625$.


\begin{figure}
\begin{center}
\includegraphics[scale=0.5]{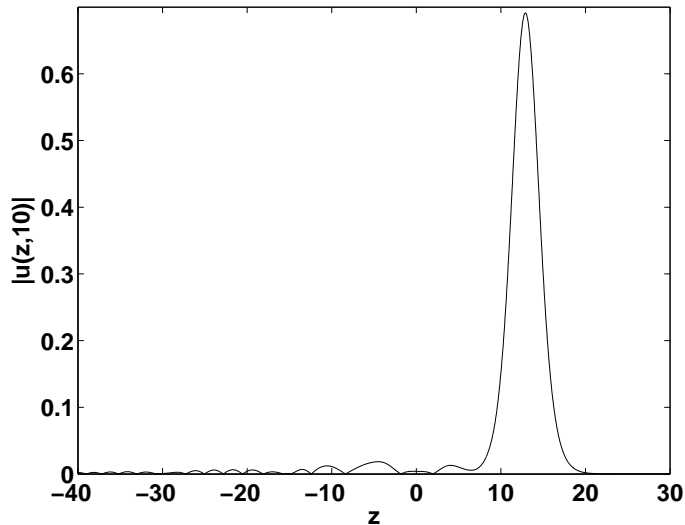}
\caption{The solution $u(z,t)$ at $t=10$ for a given disorder realization $\xi(t)$ with $D=3$ and $\epsilon=0.09$. The initial condition $u(z,0)$ is in the form of (\ref{soliton}) with $\kappa=0.5$ and $\epsilon_n = 0.64$.  }
\label{shelf}
\end{center}
\end{figure}

In the presence of perturbation, i.e., with nonzero $\epsilon$, a shelf is generated in the lee of the solitary wave and extends its range with time. Note that this interesting feature imposes major difficulties on theoretical analysis of (\ref{ekdv}). Here, we illustrate the emergence of shelf by numerically integrating (\ref{ekdv}) for a given disorder realization with $D=3$ and $\epsilon=0.09$. For the initial condition in the form of (\ref{soliton}) with $\kappa= 0.5$ and $\epsilon_n = 0.64$, figure \ref{shelf} demonstrates the solution $u(z,t)$ at $t=10$ which consists of a solitary wave and a shelf. 


\section{Numerical simulation}

We conduct Monte carlo simulations for (\ref{ekdv}) with about 100\,000 disorder realizations. Our initial condition is in the form of the solitary wave solution $u_s(z,0)$ with $\kappa= 0.5$ and $\epsilon_n = 0.64$ which corresponds to $\kappa_m=0.625$. We also assume that the disorder intensity is weak, in particular, $D=3$ and $\epsilon=0.09$. The numerical simulation is carried out until the solitary wave reaches a time $t_f = 10$, where the disorder strength becomes $D\epsilon^2t_f = 0.243$. We integrate (\ref{ekdv}) by employing a fourth-order split-step method which combines the fast Fourier transform and a fourth order Runge-Kutta scheme for the linear and the nonlinear parts of the equation, respectively \cite{agrawal}. We also introduce artificial damping near the boundaries of computational domain in order to avoid numerical artifacts resulting from the radiation emission and the use of periodic boundary conditions \cite{kms95}. The size of computational domain is taken large enough, $-100\leq z \leq 100$, so that the damping layer does not affect the major portion of solitary wave dynamics. The descritized time and spatial steps are taken as $\delta t = 0.001$ and $\delta z = 0.024$, respectively.

\begin{figure}
\begin{center}
\includegraphics[scale=0.5]{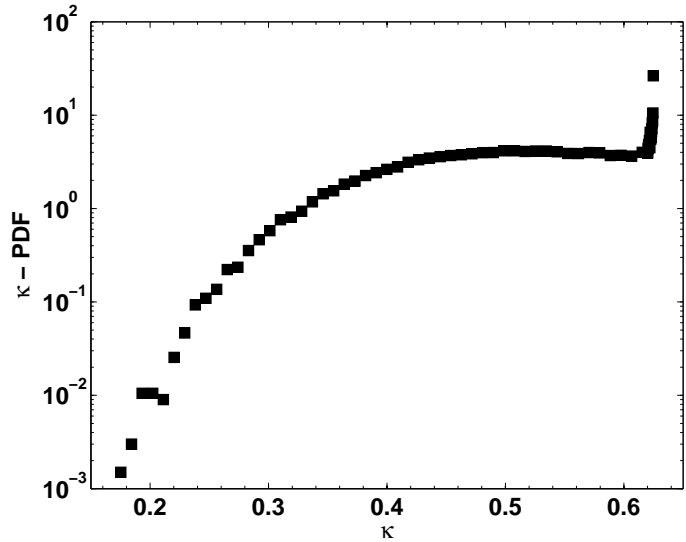}
\caption{The probability density function of $\kappa$ at $t=10$.}
\label{pdf}
\end{center}
\end{figure}

After retrieving the shape of solitary wave at the end of the evolution, we calculate the value of $\kappa$ based on (\ref{kappa}). Repeating this procedure for independent realizations of disorder, we achieve the PDF of $\kappa$ shown in figure \ref{pdf}. The result of numerical simulation clearly demonstrates that the PDF attains a divergence near $\kappa_m$. 

\begin{figure}
\begin{center}
\includegraphics[scale=0.5]{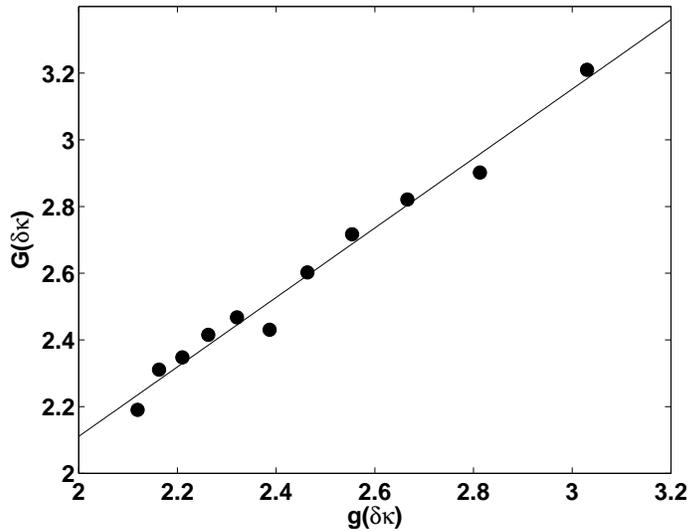}
\caption{$G(\delta\kappa)$ vs $g(\delta\kappa)$ for the same parameters used in figure \ref{pdf}. The solid line is a linear fit with slope 1.04. The circles represent the numerical result.}
\label{lin_fit}
\end{center}

\end{figure}

We now explain how to verify the divergence observed in figure \ref{pdf} is indeed loglognormal. First, it should be mentioned that a preliminary study has obtained an analytic form of $\kappa$-PDF at the first order adiabatic perturbation (see e.g., \cite{avner09} for the details of the adiabatic perturbation technique). Due to the emergence of shelf and its nature, however, a higher order perturbation theory is necessary for more exact description of the PDF. Nevertheless, the result of the first order perturbation calculation suggests that the analytic form of $\kappa$-PDF denoted by $F(\kappa)$ approximately follows (see e.g., (7) in \cite{avner09}), 
\begin{equation}
F(\kappa) \simeq \lambda_3\frac{{\rm exp}\{-\lambda_2\ln^2[\lambda_1{\rm arctanh}(\kappa/\kappa_m)]\}}{\kappa_m(1-\kappa^2/\kappa_m^2){\rm arctanh}(\kappa/\kappa_m)}, \label{ana_pdf}
\end{equation}
for $0\leq \kappa < \kappa_m$ and $F(\kappa)=0$ elsewhere. Here, $\lambda_1,\lambda_2,\lambda_3$ are some constants related to the parameters $D,\epsilon$, and the total evolution time $t_f$. This expression indicates that the loglognormal divergence can be observed in the vicinity of $\kappa_m$. Specifically, (\ref{ana_pdf}) yields an asymptotic expression of $F(\kappa)$ near $\kappa_m$,
\begin{equation}
F(\kappa)|_{\kappa \lesssim \kappa_m} \simeq \lambda_3\frac{{\rm exp}\{-\lambda_2\ln^2[-\frac{1}{2}\lambda_1\ln[\delta \kappa/(2\kappa_m)]]\}}{\delta \kappa |\ln[\delta\kappa/(2\kappa_m)]|}, \label{ana_pdf_app}
\end{equation}
where $\delta \kappa = \kappa_m-\kappa$ and $0\leq \delta \kappa/\kappa_m \ll 1$. We find that our numerically obtained PDF best fits this asymptotic expression with the constants $\lambda_1=1.7696,\lambda_2=0.8342,\lambda_3=1.0983$. For more clear demonstration of the asymptotic behaviour of numerically obtained PDF, we employ a method which allows us to map the small neighborhood of $\kappa_m$ into a wider range. Following the procedure applied to analyze the PDFs for CQNLSE and DCQNLSE \cite{avner09}, we rewrite (\ref{ana_pdf_app}) as
\begin{equation}
-\ln\left[\frac{1}{\lambda_3}\delta \kappa |\ln[\delta\kappa/(2\kappa_m)]|F(\kappa)\right] 
\simeq \lambda_2\ln^2\left[-\frac{1}{2}\lambda_1\ln[\delta \kappa/(2\kappa_m)] \right]. 
\end{equation} 
We now define $G(\delta\kappa)$ and $g(\delta\kappa)$ as
\begin{equation}
G(\delta\kappa) = -\ln\left[\frac{1}{\lambda_3}\delta \kappa |\ln[\delta\kappa/(2\kappa_m)]|F(\kappa)\right],\label{G}
\end{equation}
and
\begin{equation}
g(\delta\kappa) = \lambda_2\ln^2\left[-\frac{1}{2}\lambda_1\ln[\delta \kappa/(2\kappa_m)] \right].
\end{equation}
Notice that if the numerically obtained PDF can be described by the rhs of (\ref{ana_pdf_app}), the graph of $G$ vs $g$ is a straight line with a slope close to 1. By plugging the numerically obtained PDF data into $F(\kappa)$ in (\ref{G}), we calculate $G(\delta\kappa)$ and present the graph of $G(\delta\kappa)$ versus $g(\delta\kappa)$ in figure \ref{lin_fit}. The graph demonstrates that our numerically obtained data lies on a straight line with a slope 1.04. This result concludes that the numerically obtained PDF of $\kappa$ exhibits a loglognormal divergence in the vicinity of $\kappa_m$.


\section{Conclusion}

We investigated numerically the evolution of flat-top solitary waves by the extended Korteweg-de Vries equation. Taking into account the disorder in linear growth/damping coefficient, one of the most common disorder forms in the nonlinear wave system, we showed that the PDF of $\kappa$ which characterizes the solitary wave amplitude, exhibits loglognormal divergence near the maximum value of $\kappa$. We expect this loglognormal divergence in the vicinity of maximum possible value of $\kappa$ is mainly related to the fact that the solitary wave forms the table-top shape as the amplitude of wave approaches its maximum. The eKdV equation fundamentally differs from NLS type of equations considered earlier, however, it shows similar statistical behaviour for flat-top solitary waves. This phenomena suggests that the loglognormal divergence can be a general consequence associated with flat-top solitary waves in the presence of weak dissipative disorder.

{\ack
The author would like to thank A. Peleg for valuable comments and useful discussions.
}

\end{document}